# A Comparison of Various Turbulence Models for Analysis of Fluid Microjet Injection into the Boundary Layer over a Flat Surface

Mohammad Javad Pour Razzaghi[1], Seyed Mojtaba Rezaei Sani[2], Yasin Masoumi[3], Guoping Huan [1]*

[1]College of Energy and Power Engineering, Nanjing University of Aeronautics and Astronautics, Nanjing, 210016, PR China

[2]Department of Physics, North Tehran Branch, Islamic Azad University, Tehran 16511-53311, Iran

[3]Acoustics Research Laboratory, Center of Excellence in Experimental Solid Mechanics and Dynamics, School of Mechanical Engineering, Iran University of Science and Technology, Narmak, Tehran 16846-13114, Iran

*[hgp@nuaa.edu.cn](hgp@nuaa.edu.cn)

**Abstract**  The present work studied various models for predicting turbulence in the problem of injecting a fluid microjet into the boundary layer of a turbulent flow. For this purpose, the one-equation Spalart-Allmaras (SA), two-equation $k$-$\varepsilon$ and $k$-$\omega$, multi-equation transition $k$-$k_L$-$\omega$, transition shear stress transport (SST), and Reynolds stress models were used for solving the steady flow. Moreover, the transition SST, scale-adaptive simulation (SAS), and detached eddy simulation (DES) models were used for the transient flow. A comparison of the results indicated that the steady solution methods performed sufficiently well for this problem. Furthermore, it was found that the four-equation transition SST model was the most accurate method for the prediction of turbulence in this problem. This model predicted the velocity along the $x$-axis in near- and far-jet locations with about 1% and 5% errors, respectively. It also outperformed the other methods in predicting Reynolds stresses, especially at the center (with an about 5% error).

______________________________________________________________________

## 1. Introduction

Regarding vast applications in engineering, the problem of microjet injection into a fluid continuum has attracted many researchers [1-4] using different methods to study the problem. Increased computational power, on the one hand, and costly experimental methods, on the other, has caused numerical methods to be considered valuable analysis tools. A critical challenge in adopting these methods for studying turbulent flows is the selection of a suitable turbulence prediction method. As no specific method has been proven so far to be fully successful for this problem, and some of the methods are computationally expensive to be implemented, various methods, each with their own pros and cons, are commonly proposed for different types of flows [3, 5]. Any of these methods can outperform the others when applied under their right conditions.

The Spalart-Allmaras (SA) model is a one-equation Reynolds-averaged Navier-Stokes (RANS)-type turbulence model which can be used with 2D and 3D structured and unstructured computational grids [6]. Solutions from this model converge relatively fast for steady problems. Moreover, wall boundary conditions in this model are applied with no significant difficulty. The SA model can be calibrated for the boundary layer over a flat surface in aerodynamics problems and predicts the pressure gradient within the boundary layer [6]. It also provides satisfactory solutions for boundary layer flows, especially those included with an inverse pressure gradient. The model yields satisfactory solutions nearly up to $Y^+ = 30$ [7]. It calculates the deformation tensor through vorticity-based or strain/vorticity-based production methods. To improve the behavior of the SA model in problems with curved flow lines, curvature correction can also be incorporated into the calculations [7, 8]. Since turbulent kinetic energy (TKE) is not defined in this turbulence model, the Reynolds stress cannot be calculated.

In order to enhance the efficiency and accuracy of turbulence models, various multi-equation models have been proposed. Due to their satisfactory accuracy and speed, two-equation models are more commonly used than other multi-equation models. The *k-ε* family of turbulence models is the most widely used and most practical group of two-equation models [9]. In a *k-ε* model, two transfer equations for the turbulent kinetic energy, *k*, and the vortex dissipation rate, *ε*, are solved coupled with RANS equations. The major difference in the results from the members of this family of models can be associated with the calculation method of vortex dissipation rate, the constants used in the models, and the turbulent Prandtl number in the *k* and *ε* equations.

Being economical and stable, and as an important member of its family, the empirical standard *k-ε* model is widely used and performs well in the case of high-Reynolds-number flows [10]. It is used in fully turbulent flows only, as it neglects molecular viscosity effects. Moreover, it does not perform as well when applied to rapidly separated or highly rotational flows. The renormalization group (RNG) *k-ε* turbulent model is derived from the *k-ε* family, whose equations and constants are analytically determined through the RNG theory [11]. This model differs from the standard model in the calculation of the *ε* term. It is used in problems involving boundary-layer separation, vortex shedding, and shear flows with average rotation rates. The realizable *k-ε* turbulence model outperforms the other members of the *k-ε* family when the flow contains an inverse gradient or experiences separation [12]. This model can be used in problems involving rotational flows or high strain rates. However, the accuracy of the model drops near walls and in the case of low-Reynolds-number flows [12]. Furthermore, it takes advantage of wall functions to calculate flow properties near walls. Accordingly, non-equilibrium wall functions are preferred in the case of high-Reynolds-number flow, where the flow through the viscous sublayer cannot be solved by allocating computational cells within the sublayer. Moreover, standard wall functions are used for simple shear flows. Non-equilibrium wall functions are suitable for flows involving low separation, those sticking back to the surface after separation, those having an intense pressure gradient, and flows dispersing over the surface. On the other hand, these functions are not recommended in cases of low-Reynolds-number flows, blowing/suction simulations, and pressure gradients leading to boundary layer separation [12].

In simulating flows with a velocity drop or separation caused by inverse pressure gradients, the two-equation *k-ω* model is better and more widely used than the *k-ε* model for near-wall turbulence calculations [13]. The standard *k-ω* model can improve near-wall (viscous sublayer) calculations for low-Reynolds-number numbers without using wall functions. This model performs well in the analysis of boundary layers in wall-bounded flows, free shear flows, flows with inverse pressure gradients, and low-Reynolds-number flows [14]. However, the model is highly sensitive to free-flow conditions. Depending on the value of *ω* at the inlet, it can lead to different results from the same simulation. To resolve this problem, the baseline (BSL) *k-ω* model is used [15]. This model takes advantage of the properties of the *k-ω* model for near-wall regions and those of the *k-ε* model for far-wall regions.

In simulations with an emphasis on the boundary layer or those with adverse pressure gradients, the shear stress transport (SST) *k-ω* model can be used. This method, in turn, is faced with difficulties for high-Reynolds-number flows [13]. Furthermore, low-Reynolds-number and shear flow corrections are applied to the model when it is used for the prediction of low-Reynolds-number and shear flows, respectively. Also, production limiters are used to prevent the unjustified increase in the TKE at stagnation points [13]. The transition *k-$k_L$-ω* turbulence model is used to predict boundary layer development and calculate the start of the transition. This model can be effectively used for the transition from a laminar boundary layer to a turbulent boundary layer. However, it is computationally more expensive than two-equation models [16]. The transition SST turbulence model can be used for solving problems where the boundary layer is mostly laminar and for detecting the transition zone. Nevertheless, it requires a fine grid, which may prolong computations [17, 18].

The Reynolds stress model (RSM) of turbulence can be used in cyclone flows, highly rotational flows inside combustion chambers, and highly swirling flows [19]. This model uses the linear pressure-strain model to calculate the pressure strain. The stress-$\omega$ and stress-BSL models are also used to calculate the specific dissipation rate. While the former is used in calculations involving curved surfaces and rotational flows, the latter can reduce sensitivity to free stream. The wall boundary condition and the wall reflection effects from the $k$ equation can also be used in the stress-BSL model [20]. The scale-adaptive simulation (SAS) model is used to improve the solution method for RANS equations. Provided that a grid of proper quality is used, this model can lead to results similar to those of large eddy simulation (LES) and detached eddy simulation (DES) methods [21]. It can also be used in aerodynamics problems involving strong flow separation. Moreover, this method can model micro vortices [22]. Unsteady RANS (URANS) models are utilized in DES for solving boundary layers, whereas LES is employed to simulate flow separation zones. There are various RANS models for DES methods [23]. DES models were essentially developed for wall-bounded high-Reynolds-number flows, as there are high computational costs (CPU and time) associated with LES methods when solving a near-wall turbulent flow field. Combined DES models differ from LES models only in using RANS models to solve the near-wall flow field problems [23].

In 2005, Park et al. [24] improved the solutions of two-equation turbulence models by incorporating limiters in the model and analyzed the effects of limiters on various flows such as the simple shear flow, supersonic compression ramp flow, and supersonic base flow. Four years later, Durbin [25] studied the effects of limiters and wall treatments on the existing turbulence models and analyzed the $k$-$\omega$ model and the effects of these limiters on predicting the results from this model. Wang et al. [26] in 2012 used limiters in turbulence models to analyze heat transfer in a flow over a flat surface. They compared two $k$-$\varepsilon$ models with and without limiters and analyzed the effects of various coefficients of the limiters on the final results by varying the model parameters. Moreover, they found that the limiters improved the calculation of some of these parameters. A study by Li et al. [27] considered various turbulence models for analyzing air pressure in a building complex. They focused on and discussed the advantages of the SA, $k$-$\varepsilon$, and $k$-$\omega$ models. In another study in 2015, Králik et al. [28] analyzed turbulence models of flow over a quarter-circular object. They studied the flow and compared $k$-$\varepsilon$, $k$-$\omega$, LES, and SAS models. A year later, Ramdlan et al. [29] analyzed the $k$-$\varepsilon$, RNG $k$-$\varepsilon$, and Reynolds stress turbulence models of a wind tunnel and compared the turbulence parameters of the models. Wienand et al. [30] in 2017 used two-equation turbulence models to study an impinging jet on a flat plate. They used SST models and a limiter to analyze the flow by varying the jet-to-plate distance. Recently, John et al. [31] numerically analyzed the $k$-$\varepsilon$, $k$-$\omega$, RSM, SST $k$-$\omega$, and LES models of a wind tunnel and studied the flow velocity and turbulence parameters in different cross-sections of the tunnel. In another recent study, Zhang et al. [32] analyzed the $k$-$R$ turbulence model for wall-bounded flows. They compared the results from this model and those from the DNS, $k$-$\varepsilon$, and SST $k$-$\omega$ models and found that all of them predicted almost similar results. However, the results from their proposed model followed the experimental results more closely under specific conditions.

The present work has analyzed the effects of fluid microjet on a turbulent boundary layer. For this purpose, a microjet was injected into the flow to create a vortex, which caused the energy to be injected from the upper layers of the flow down into the boundary layer. This caused the flow to stick to the surface and delayed flow separation. As discussed above, proper selection of a turbulence model is essential for accurately modeling a turbulent flow. A review of previous researches revealed that various turbulence models should be analyzed to select which one is the best for the problem in question. To this end, different one-, two-, and multi-equation turbulence models were analyzed in steady and transient conditions, and ultimately the pros and cons of the best models were discussed. It was shown that it is possible to take advantage of the benefits of various models by considering the final required output.

The remaining of this paper is organized as follows. Section 2 is devoted to the details of modeling, governing equations, and verification of results. Section 3 discusses the obtained results for different turbulence models. Finally, we present our concluding remarks on the comparison of results of the considered turbulence models.

## 2. Numerical Modeling

### 2.1 Modeling Details

To model the turbulent flow, the Ansys FLUENT software [33] was employed. The geometry of the experimental wind tunnel was modeled as shown in Fig. 1a, where the dimensions, the origin of the coordinate system, and the boundary conditions are specified. The model dimensions in the $x$- and $z$-directions were 3.5 and 0.5 m, respectively. The values of $y$ at the inlet and outlet ends of the tunnel were 0.130 and 0.231 m, respectively. The microjet was located 1.130 m from the inlet inside the tunnel. It was 0.001 m in diameter and located 0.005 m above the line $z = 0$ m at the bottom of the channel. Fig. 1b shows the orientation of the microjet relative to the coordinate axes. The fluid considered was the air with viscosity $\mu = 1.81 \times 10^{-5}\ kg/s\ m$ and density $\rho = 1.12\ kg/m^3$. The inflow velocity was 15 m/s, and its turbulence intensity was 2%. The ratio of turbulent to molecular viscosity at the inlet and outlet boundaries was set to 5% and 10%, respectively. According to previous studies [1, 34, 35], a 30° pitch angle and a 60° skew angle were selected, and the microjet was decided to inject at a rate of 60 m/s to yield the best performance. Tab. 1 presents the geometrical definitions of the problem as well as its inputs and outputs.

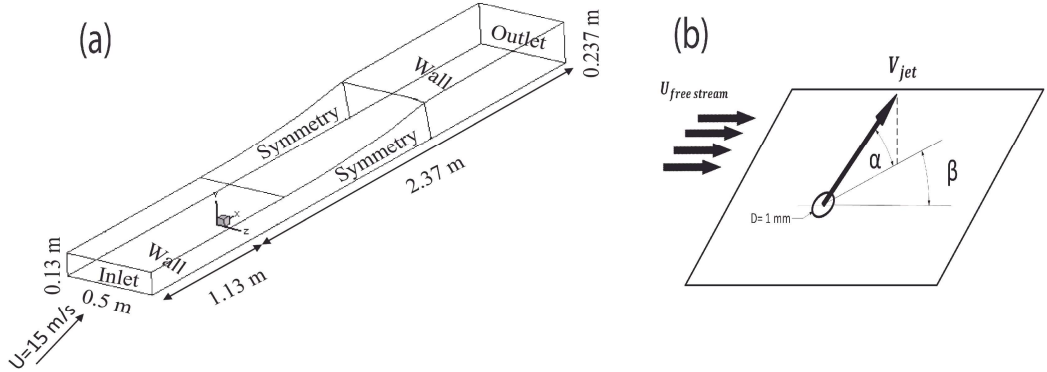

Figure 1. (a) Dimensions and boundary conditions of the computational domain. (b) Orientation of the microjet relative to the coordinate axes.

Table 1. Parameters values and their definitions.

| Parameter | Definition | Parameter | Definition |
|---|---|---|---|
| $L$ = 3.5 m | Total surface length | $D$ = 0.001 m | Microjet diameter |
| $\rho = 1.12\ kg/m^3$ | Fluid density | $h$ = 0.005 m | Distance of the microjet from the center of the surface |
| $\mu_t/\mu$ = 5% – 10% | Ratio of turbulent to molecular viscosity | $\alpha$ = 30° | Pitch angle of jet hole |
| $\mu = 1.81 \times 10^{-5}\ kg/s\ m$ | Fluid viscosity | $\beta$ = 60° | Skew angle of jet hole relative to $x$-direction |
| $U$ = 15 m/s | Inflow velocity | $V_{jet}$ = 60 m/s | Fluid microjet velocity |
| $TI$ = 2% | Turbulence intensity | | |

### 2.2 Governing Equations

Navier-Stokes equations were used to solve the surface flow. However, as the inlet flow was turbulent, their direct modeling requires considerable time and computational costs. Therefore, the governing equations were expanded into RANS equations and solved numerically. The equations were discretized using the finite volume method, and velocity and pressure fields were coupled through the SIMPLE algorithm [36]. The momentum and pressure equations were discretized using the second-order upwind scheme [9]. Turbulent flow modeling was carried out using the SST $k$-$\omega$ model and first-order upwind scheme was used for discretization of turbulence dissipation rate and turbulent kinetic energy. So, to solve the RANS equations, the continuity and momentum equations in incompressible flow are used as in Eqs. 1 and 2, respectively [37]

$$\rho \frac{\partial}{\partial t}\mathbf{v} + \rho(\mathbf{v}.\boldsymbol{\nabla})\,\mathbf{v} = \boldsymbol{\nabla}.(\boldsymbol{\sigma}) + \mathbf{b}_\mathrm{f}, \quad (1)$$

$$\boldsymbol{\nabla}.\mathbf{v} = 0, \qquad (2)$$

where $\boldsymbol{\sigma}$ is fluid stress tensor and is equal to $\boldsymbol{\sigma} = -p\mathbf{I} + (\mu + \mu_t)[\boldsymbol{\nabla}\mathbf{v} + (\boldsymbol{\nabla}\mathbf{v})^\mathrm{T}]$, $p$, $\mu$, $\mu_t$, $\rho$, $\mathbf{v}$, $\mathbf{b}_\mathrm{f}$, and $\mathbf{I}$ are static pressure, dynamic viscosity, turbulence viscosity, fluid density, fluid velocity, body force, and the identity tensor, respectively.

## 2.3 Verification of Results

As stated earlier, this paper numerically analyzes the effects of a microjet on a turbulent surface flow. To this end, grid-independent tests were carried out by evaluating the numerical results. The results from the numerical model and those from the experimental wind tunnel were also compared, and the validity of the numerical modeling was verified. Fig. 2 depicts the considered geometry for which a hexahedral (HEX) mesh model is used for grid generation. The mesh is finer in regions with more important and intensified flow variations, such as the lower and the inlet boundaries of the domain.

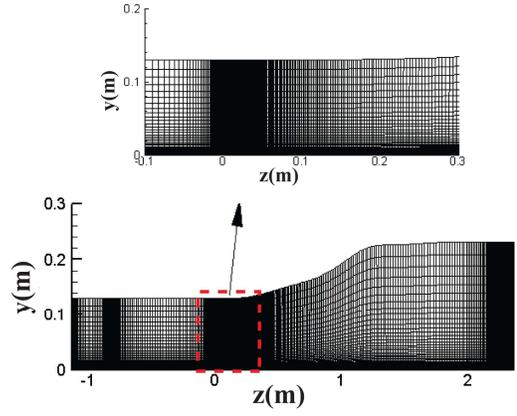

Figure 2. The mesh grid used for the computational domain.

To perform the grid-independent test for the case without a microjet, a comparison was made between three mesh densities: coarse, normal, and fine grids having 1024128, 1922544, and 3894000 cells, respectively. Fig. 3a represents the $x$-components of the velocity for a line parallel to the $y$-axis at $x/L = 0.1714$. According to the figure, the $x$-component of the velocity is zero near the wall and gradually increases at points farther from the bottom, up to a distance of about $y/\delta = 1$, after which it remains constant. A negligible discrepancy of about 1% is observed between the results for different mesh densities in regions with more intense variations. Fig. 3b compares the corresponding calculated pressure coefficients at the bottom surface of the wind tunnel. It is observed that the pressure decreases almost linearly in the front portion of the plane. According to Bernoulli's equation, moving toward the regions with a larger distance between the walls, one can see a pressure increase due to lower velocities in the longitudinal direction. There is also a good agreement between the results from the different mesh densities. Slight discrepancies between the results from the coarse grid and those from the other two mesh densities are seen. The largest difference occurred at $x = 1.3$ m and was below 1%. The results for the two turbulence parameters (Reynolds stresses) $u'u'$ and $u'v'$ at $x/L = 0.1714$ from the origin along the $y$-axis are shown in Figs. 3c and d for comparison. The figures show that the absolute values of these parameters increased sharply from zero at the lower boundary. Next, the values returned to zero in regions closer to the middle of the tunnel, where there was almost no velocity variation then. The difference in the results at points near the lower boundary was less than 1%. The results indicate that the three mesh densities could provide reasonable accuracy for the present problem. However, adding a microjet resulted in more substantial velocity variations near the boundary and led to larger errors in the case of the coarse grid. Therefore, the normal grid density was chosen for the next stages of the analysis.

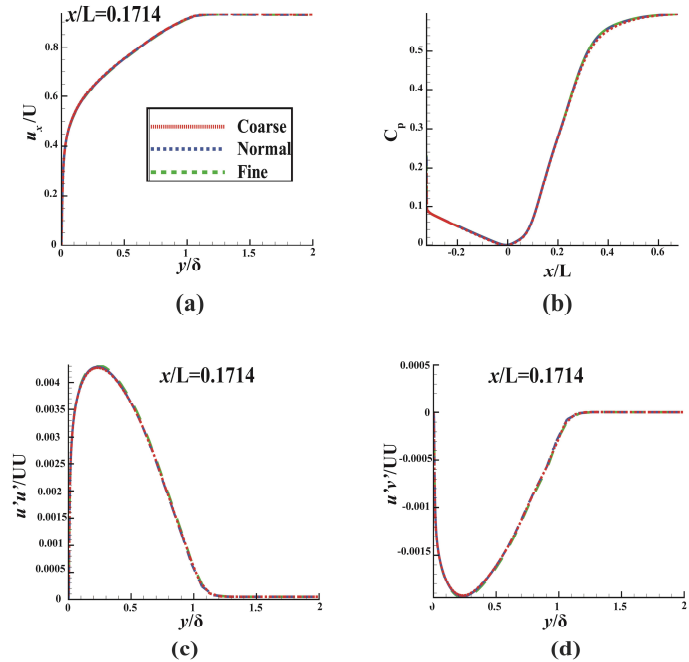

Figure 3. Comparison of (a) $u_x$, (b) $C_p$, (c) $u'u'$, and (d) $u'v'$ for coarse, normal, and fine mesh densities at the lower wall.

The values for $Y^+$ over the lower walls of the tunnel are shown in Fig. 4. It is observed that the value of $Y^+$ at any point on the lower wall was below the permissible limit (about 300 [38]). This issue confirms the mesh was acceptable near the walls for a turbulent flow.

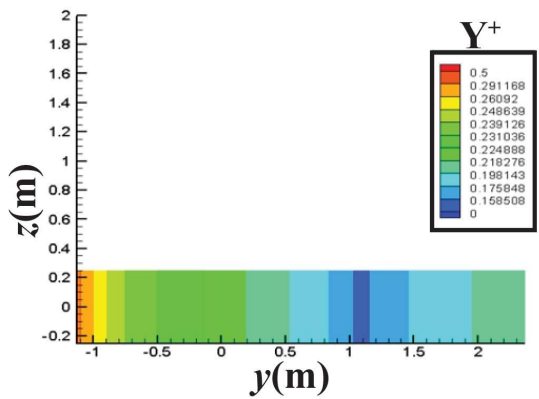

Figure 4. Values of $Y^+$ at the lower wall of the wind tunnel.

The verification experiment was completed in a low-speed closed-circuit boundary layer wind tunnel. An adverse pressure gradient turbulent boundary layer was measured with and without the microjet. Hot-wire technique was applied to measure the flow velocities and Reynolds stresses. The wire anemometer used was FML constant temperature anemometers operating at a resistance ratio of 1.8. The single wire probe was a DISA 55 POS probe. The cross wire probe was DOSA 55-p-51 dual-sensor hot wire probe. The probes used 2.5us platinum-coated tungsten wire which was copper plated and then etched for an active length to diameter ratio of 250. The data acquisition was performed with a586PC and National Instruments PCI-MID-16 Board (12-bit A/D). The single wire signals were sampled at 1 kHz for 20 seconds giving 20000 samples per $y$ position in the profiles. At each measurement position, 6000 samples from each wire of cross wire probe were acquired using the external simultaneous sample and held at a frequency of 60 Hz. Because there was a trip at 53.3 mm downstream of the leading edge, the flow of the boundary layer was well developed turbulent boundary layer. The two-axis traverse system has a resolution of 0.0032 mm in the vertical direction and 0.0064 mm in the spanwise direction. The spatial resolution of the hot-wire measurements was 0.65 mm × 1.0 mm ($y \times z$). The longitudinal pressure gradient was measured using wall taps. The static pressure data was measured by the Celesco P7D diaphragm-type pressure transducer with a range of ±0.1 psi [39, 40].

According to the uncertainty analysis of Anderson and Eaton [41], the mean velocity had an uncertainty of 3% of the local streamwise velocity. The normal Reynolds stress components revealed an uncertainty of 5% of the local value of $\overline{u'^2}$. Also, the shear stress showed an uncertainty of 10% of the local value of $\overline{u'v'}$.

To verify the numerical modeling, Figs. 5 and 6 present a comparison of the experimental and numerical pressure coefficients and Reynolds stresses. On the lower wall of the wind tunnel, good agreement is observed between the experimental and numerical pressure coefficients, as shown in Fig. 5. A 3% difference is observed between the numerical and experimental results due to the geometric inconsistencies (in the upper curve) of the actual tunnel and the (not wholly symmetrical) wall boundary conditions.

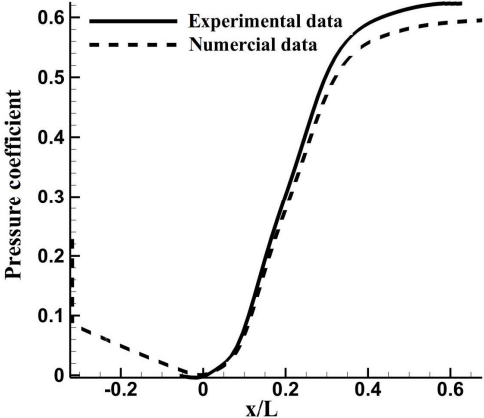

Figure 5. Comparison of the numerical and experimental pressure coefficients.

Fig. 6 shows the numerical and experimental Reynolds stresses at different cross-sections of the turbulent boundary layer. The results indicate that there is good agreement between the numerical and experimental values. However, there is a slight difference in some cases. While many cases showed a difference lower than 2% between the experimental and numerical values, the difference grew to 9% for some other cases. This might be the result of probable experimental errors due to measurement sensitivities. However, the numerical results were concluded to be reasonably accurate.

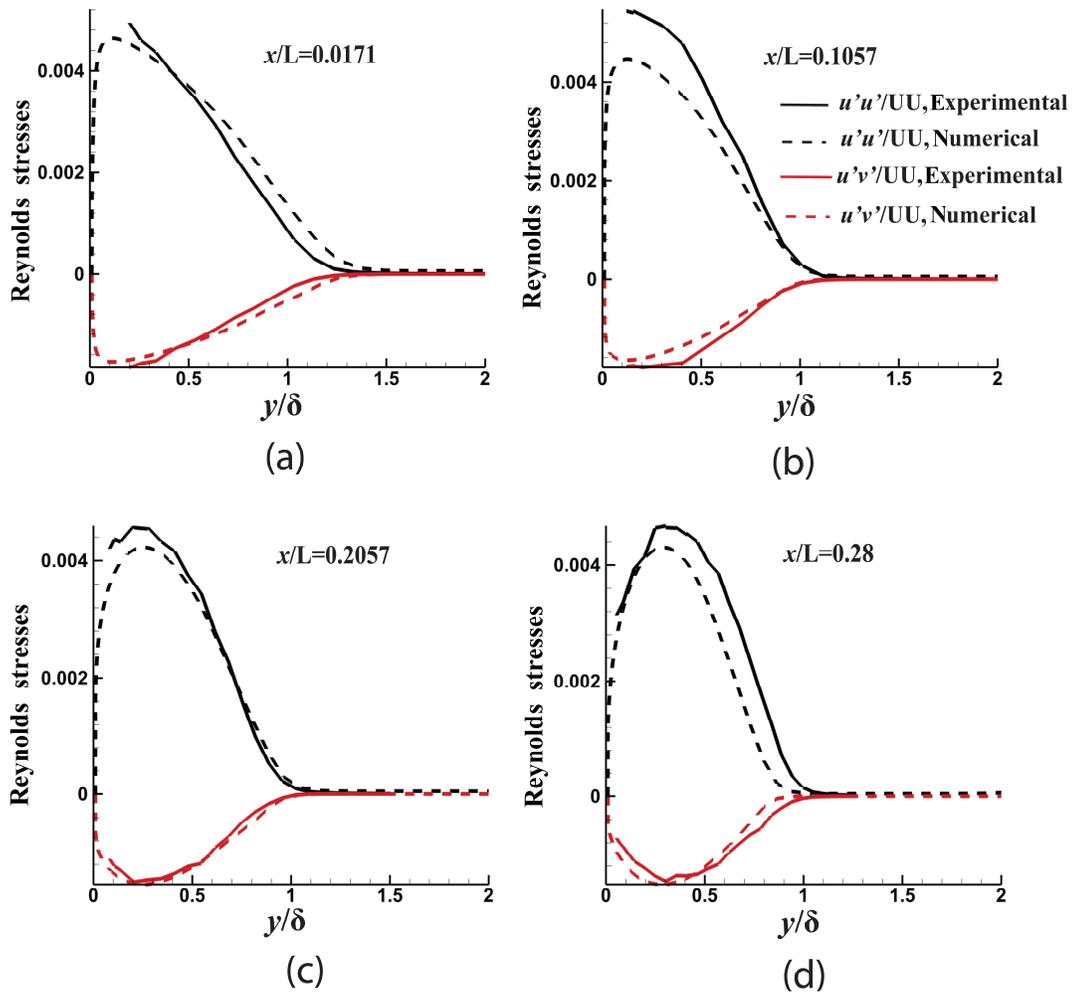

Figure 6. Comparison of numerical and experimental Reynolds stresses u'u' and u'v': (a) $x/L = 0.0171$, (b) $x/L = 0.1057$, (c) $x/L = 0.2057$, and (d) $x/L = 0.28$.

Fig. 7 compares the experimental and numerical results for a surface flow with a microjet. The comparison was made at 9 cm from the microjet by considering u/U for VRs = 1, 2, and 4, corresponding to percentage errors of 1%, 3%, and 5%, respectively. As the figure shows, there is good agreement between the experimental and numerical results. It was concluded that the numerical results could be used in further stages through the analysis process.

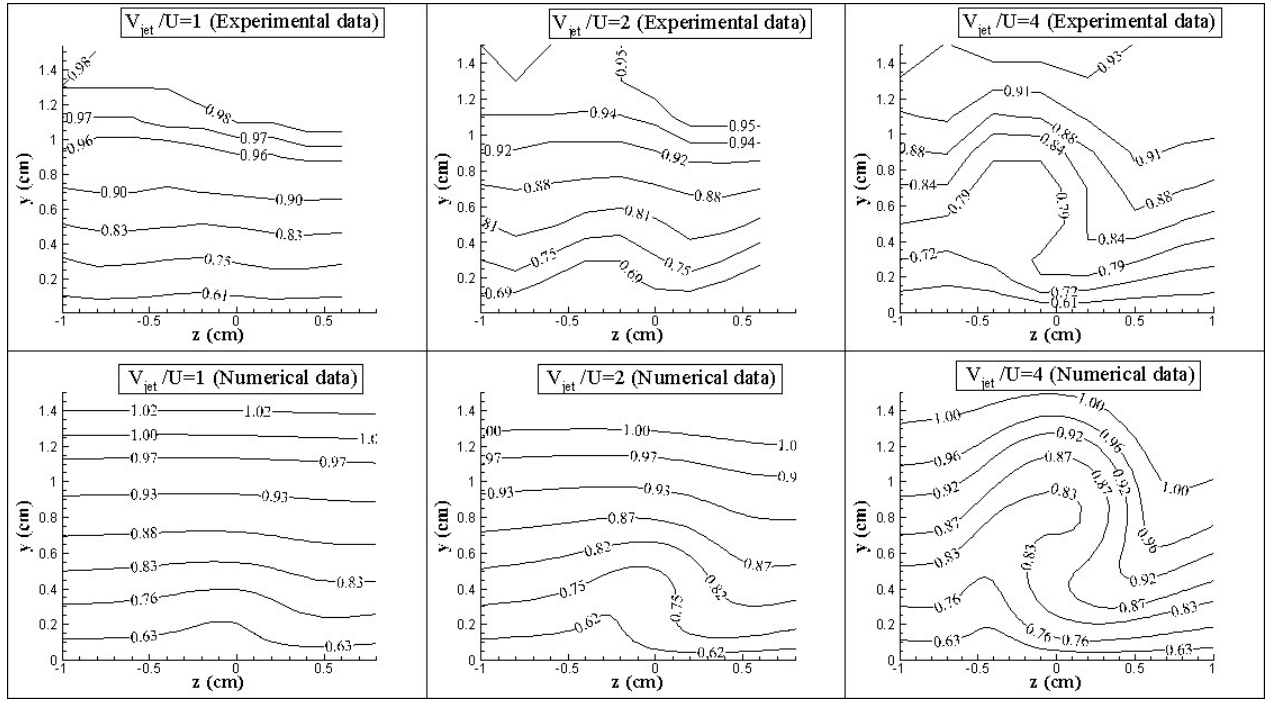

Figure 7. Comparison of the experimental (top row) and numerical (bottom row) results for $V_{jet}/U$ contours for various microjet velocities at $x = 0.09$ m.

## 3. Results and discussion

Given the above discussions, various turbulence models aiming to predict the behavior of a microjet injected into a turbulent boundary layer are analyzed in what follows. We considered one-, two-, and multi-equation models and obtained the solution for a steady flow. Furthermore, as it was impossible to use such models as SAS and DES in steady state conditions, the transient solution obtained from these models was also analyzed and compared. Velocity and Reynolds stresses were presented and compared for the vortex's center, upwash, and downwash portions of the vortex. Fig. 8 shows a schematic view of the different portions.

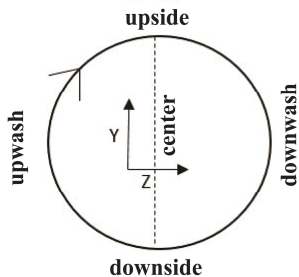

Figure 8. Definition of different regions around a vortex.

### 3.1 Spalart-Allmaras model

In order to analyze the performance of one-equation methods in modeling turbulence flows, this section discusses the one-equation SA model.

Table 2: Different cases of the one-equation SA model.

| Case 1-1 | Spalart-Allmaras (production: vorticity-based) |
|---|---|
| Case 1-2 | Spalart-Allmaras (production: strain/vorticity-based) |
| Case 1-3 | Spalart-Allmaras (production: vorticity-based), (option: curvature correction) |

Tab. 2 presents different cases of this model based on different solution methods. The results in Fig. 9 indicate that the SA model predicted $u_x$ reasonably well, especially in near-jet regions. In fact, the results from this model differed from the experimental results in regions farther from and closer to the jet by 10% and 3%, respectively. No difference was observed between the results from the different cases. At locations farther from the surface, $u_x$ values predicted by these models were lower than the corresponding values from the experiment. We conclude that if the analysis is aimed at determining variations in flow velocity, this method can be used because it is cost-effective. However, as this method only solves a single turbulence equation, it cannot predict the shear stresses; therefore, multi-equation methods are required. All SA model cases yielded similar results consistent with the experimental values. Furthermore, this model is computationally efficient and reduces calculations. However, it cannot predict such turbulence parameters as Reynolds stresses.

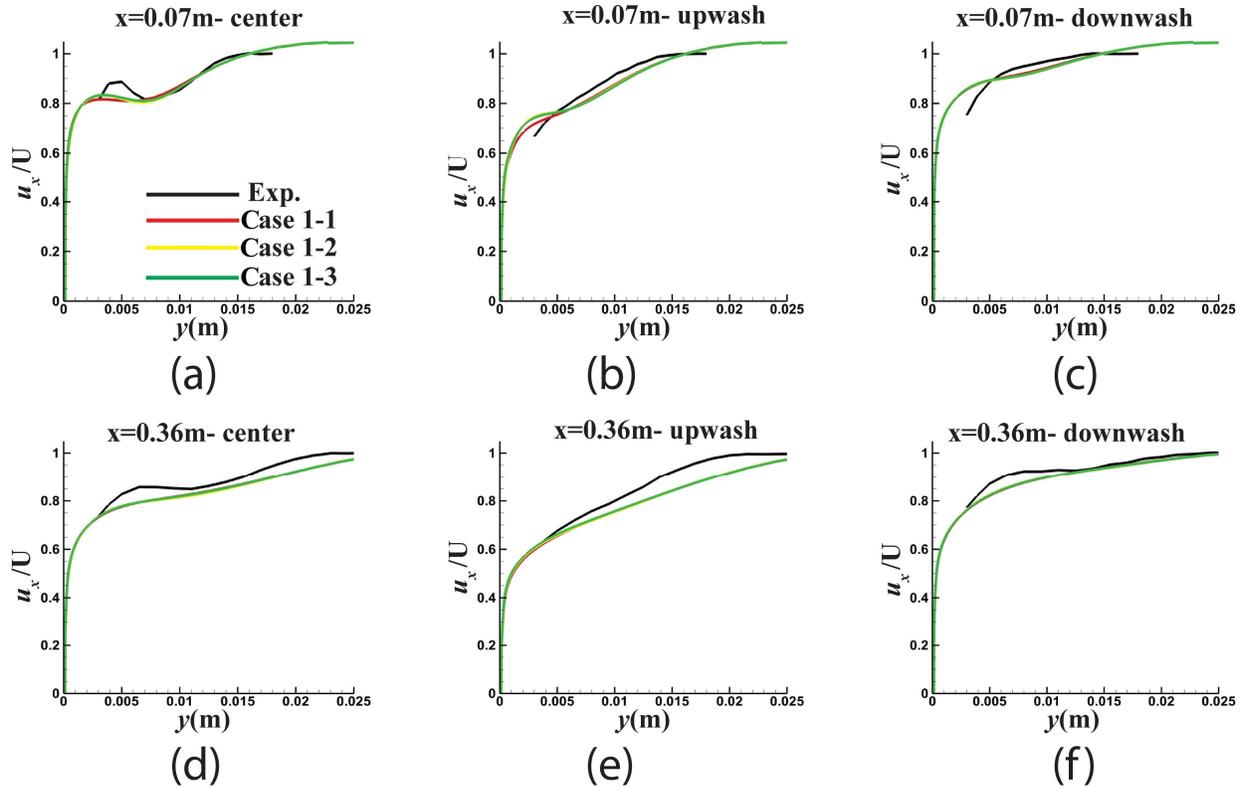

Figure 9. Velocity contours from experiment and different cases of one-equation SA model for (a) center, (b) upwash, and (c) downwash regions at $x = 0.07$ m cross-section, and for (d) center, (e) upwash, and (f) downwash regions at $x = 0.36$ m cross-section.

*3.2 k-ε model*

Tab. 3 presents various models using the two-equation $k$-$\varepsilon$ method. These models and the effects of using wall functions are analyzed in the following.

Table 3. Different cases of the two-equation $k$-$\varepsilon$ model.

|  | Model details |
| --- | --- |
| Case 2-1 | $k$-$\varepsilon$ (model: standard), (near-wall treatment: standard wall functions) |
| Case 2-2 | $k$-$\varepsilon$ (model: RNG), (near-wall treatment: standard wall functions) |
| Case 2-3 | $k$-$\varepsilon$ (model: realizable), (near-wall treatment: standard wall functions) |
| Case 2-4 | $k$-$\varepsilon$ (model: standard), (near-wall treatment: scalable wall functions) |
| Case 2-5 | $k$-$\varepsilon$ (model: standard), (near-wall treatment: Menter-Lechner) |

Fig. 10 shows the velocity along the $x$-axis predicted by the different cases. It was observed that Cases 2-4 and 2-5 predicted $u_x$ more reasonably, especially in near-jet regions. In fact, using more advanced models involving near-wall treatment improved the performance of the standard $k$-$\varepsilon$ model. The values predicted by the other models were close to each other. The difference between the $u_x$ values predicted by Case 2-5 and the corresponding values from the experiment in regions closer to and farther from the jet was about 3% and 10%, respectively.

Therefore, this method does not offer a significant advantage over the SA method in terms of velocity prediction. Overall, the results obtained from the other cases of the model were less satisfying than those from the one-equation model.

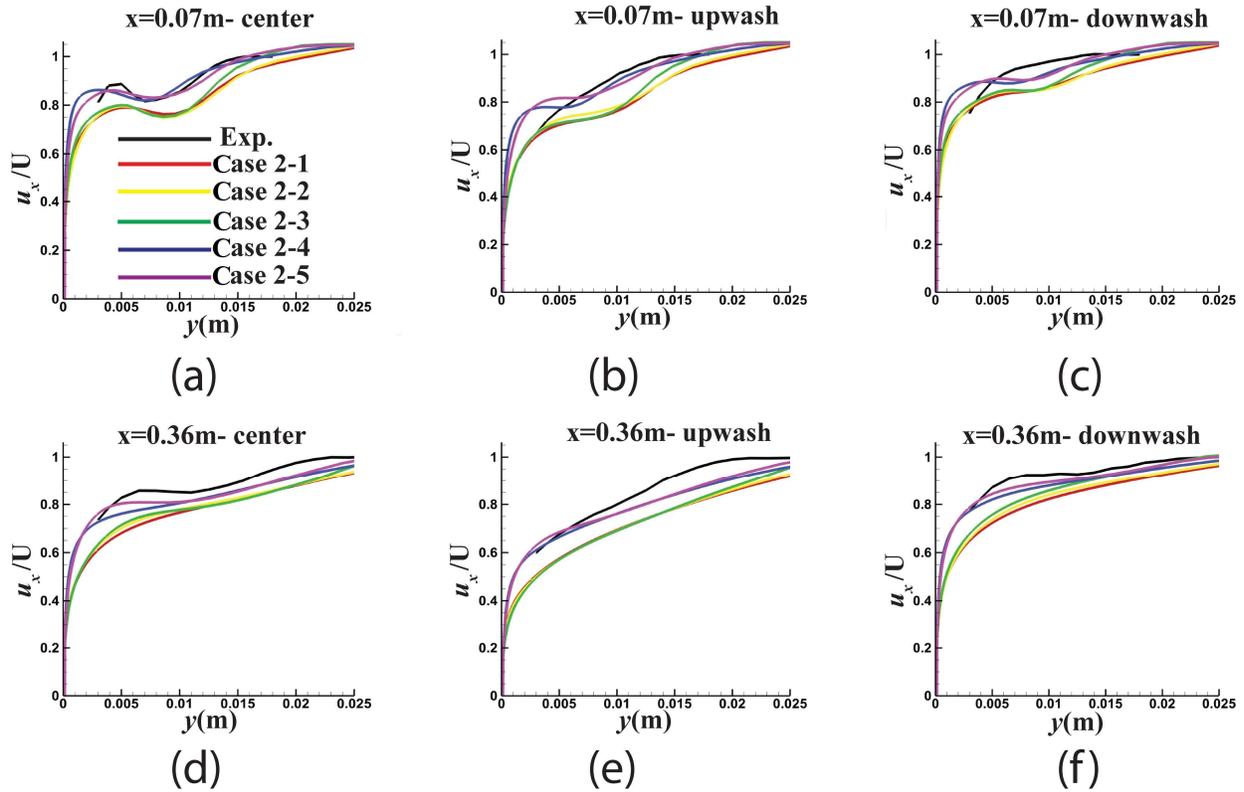

Fig. 10. Velocity curves from experiment and different cases of two-equation $k$-$\varepsilon$ model for (a) center, (b) upwash, and (c) downwash regions at $x = 0.07$ m cross-section, and for (d) center, (e) upwash, and (f) downwash regions at $x = 0.36$ m cross-section.

Figs. 11 and 12 show the Reynolds stresses $u'u'$ and $u'v'$ for different cases. These models substitute the SA method because the SA method cannot calculate these parameters. The results indicate that the models led to larger errors when used to calculate the Reynolds stresses compared to velocity. However, the best results were obtained from Case 2-5. These predictions were more accurate for far-jet regions. Furthermore, according to the figures, the results were predicted more accurately at the center. A probable source of increased error in the upwash and downwash flows is the difficulty in the experimental calculation of turbulence parameters, especially in the predictions regarding these two regions. The graphs also indicate that the results improved at regions farther from the surface.

Comparison of the results regarding the Reynolds stresses shows that $u'v'$ was predicted more reasonably than $u'u'$. The reason for the larger discrepancy associated with $u'u'$ originates from the difference in the predicted results pertaining to the TKE. The difference between the values predicted by Case 2-5 and the experimental results typically for different cross-sections reached 40% and 10% for $u'u'$ and $u'v'$, respectively. The results presented in this section indicate that Case 2-5 yielded better results compared to the other cases. The velocity values predicted by this model were similar to those predicted by the SA model. Moreover, the Reynolds stress $u'v'$ predicted by this model followed the experimental values more closely.

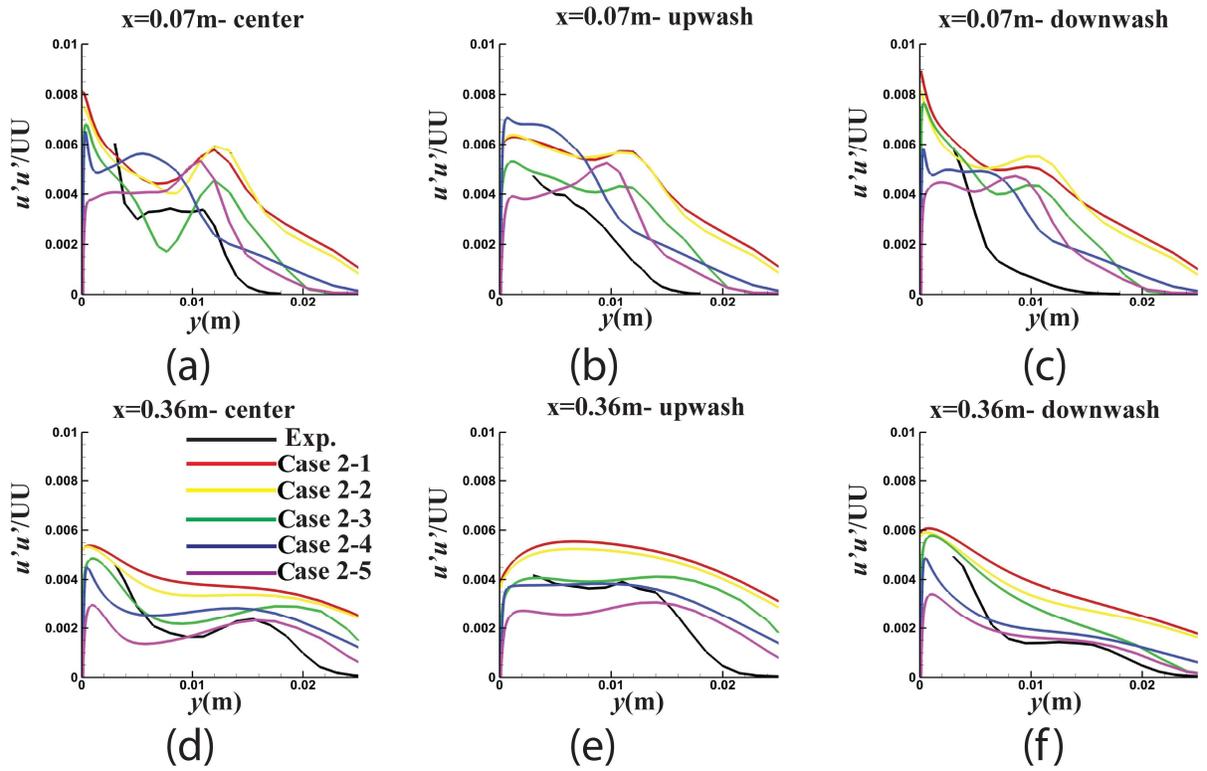

Figure 11. *u'u'* curves from experiment and different cases of two-equation *k-ε* model for (a) center, (b) upwash, and (c) downwash regions at *x* = 0.07 m cross-section, and for (d) center, (e) upwash, and (f) downwash regions at *x* = 0.36 m cross-section.

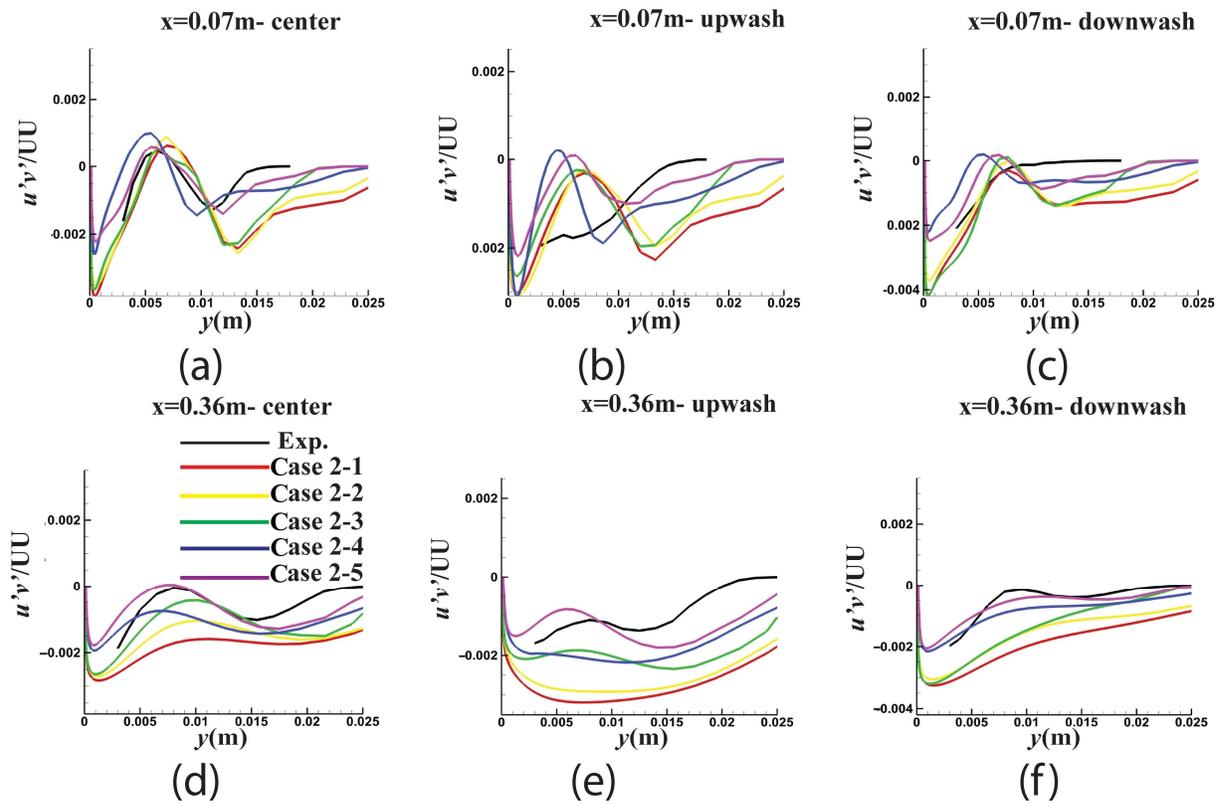

Figure 12. *u'v'* curves from experiment and different cases of two-equation *k-ε* model for (a) center, (b) upwash, and (c) downwash regions at *x* = 0.07 m cross-section, and for (d) center, (e) upwash, and (f) downwash regions at *x* = 0.36 m cross-section.

*3.3 k-ω model*

This section discusses the two-equation *k-ω* model. For this purpose, various models using this method are presented in Tab. 4 and compared in what follows.

Table 4. Different cases of the two-equation *k-ω* model.

|  | Model details |
|---|---|
| Case 3-1 | *k-ω* (model: standard), (options: shear flow corrections , production limiter) |
| Case 3-2 | *k-ω* (model: BSL), (option: production limiter) |
| Case 3-3 | *k-ω* (model: SST), (option: production limiter) |
| Case 3-4 | *k-ω* (model: SST), (options: low-Reynolds-number correction, production limiter) |
| Case 3-5 | *k-ω* (model: standard), (options: shear flow corrections, low-Reynolds-number correction, production limiter) |

As the *k-ε* method did not exhibit any significant advantage over the SA method in terms of velocity prediction, in the following, we discuss the *k-ω* method. Fig. 13 shows that, contrary to the *k-ε* method, almost all the cases of the *k-ω* method predicted similar values for the velocity. The predicted results for $u_x$ were almost similar to those from the two previous models. We can conclude that the longitudinal component of velocity does not depend significantly on the turbulence parameter due to turbulence intensity.

The Reynolds stresses obtained from the *k-ω* models are presented in Figs. 14 and 15 indicate that this method has compensated to some extent for the shortcomings of the *k-ε* method. According to the figures, there was a minor discrepancy between the results obtained from the *k-ω* method and the experimental results. Although the predictions by the turbulence models reproduced the experimental results strictly for some cases, there was a significant difference between the predictions and the experimental results for some other cases. This can be attributed to the calculation error in the vortex position. The results indicated that using different options in the standard model did not influence the results. It was also found that Case 3-4 outperformed Case 3-3 in predicting *u'u'* near the surface. For Case 3-4, the error averaged 20% for both *u'u'* and *u'v'*. We can conclude that this method exhibited better performance in predicting the TKE. The values obtained from the two-equation *k-ω* model show that Case 3-4 yielded better results. Comparison of the two-equation models also indicates that Cases 3-4 and 3-5 are preferred for calculating *u'u'* and *u'v'*, respectively.

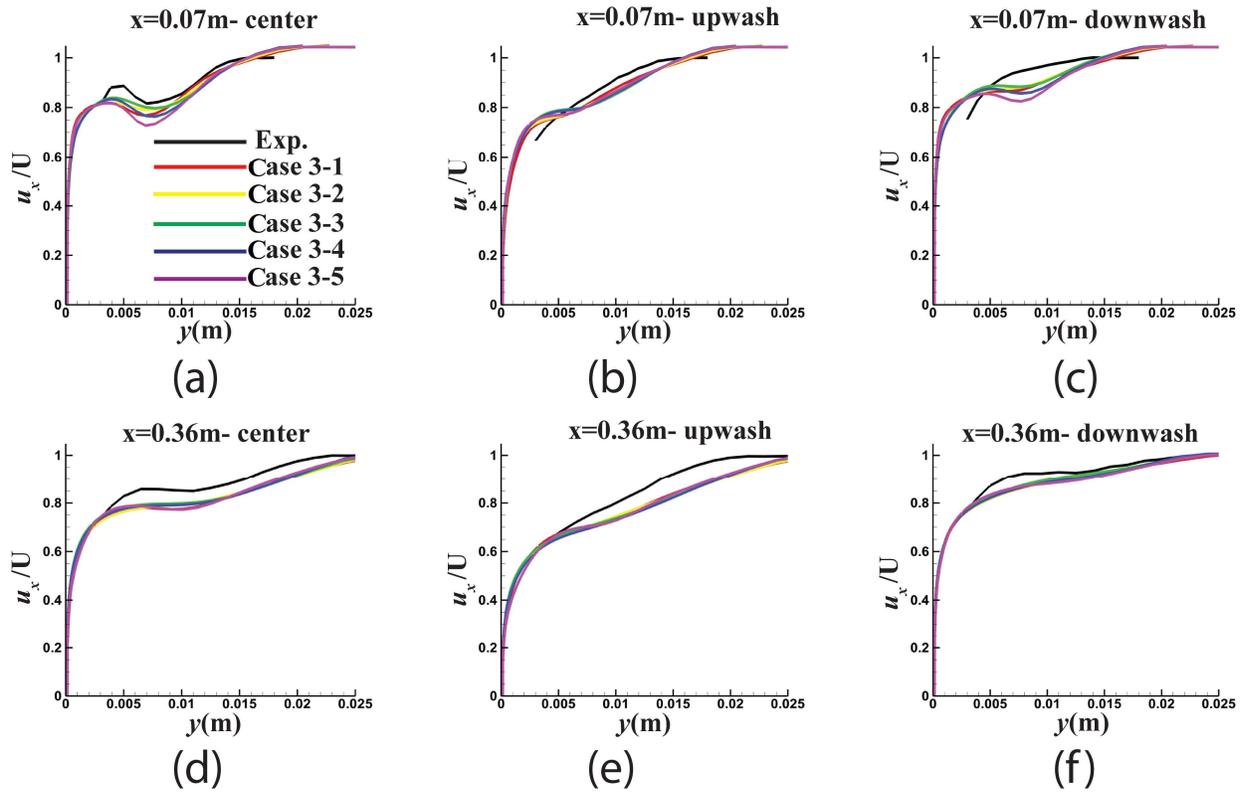

Figure 13. Velocity curves from experiment and different cases of two-equation $k$-$\omega$ model for (a) center, (b) upwash, and (c) downwash regions at $x = 0.07$ m cross-section, and for (d) center, (e) upwash, and (f) downwash regions at $x = 0.36$ m cross-section.

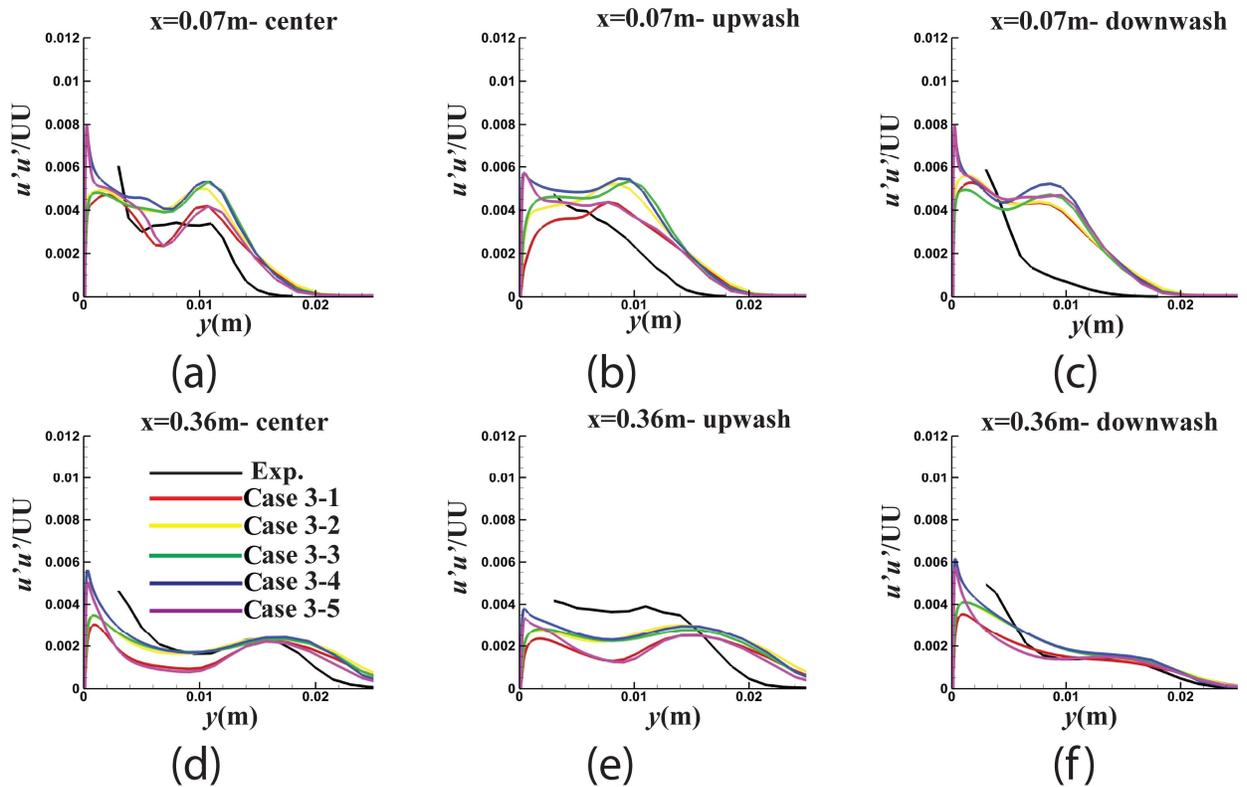

Figure 14. $u'u'$ curves from experiment and different cases of two-equation $k$-$\omega$ model for (a) center, (b) upwash, and (c) downwash regions at $x = 0.07$ m cross-section, and for (d) center, (e) upwash, and (f) downwash regions at $x = 0.36$ m cross-section.

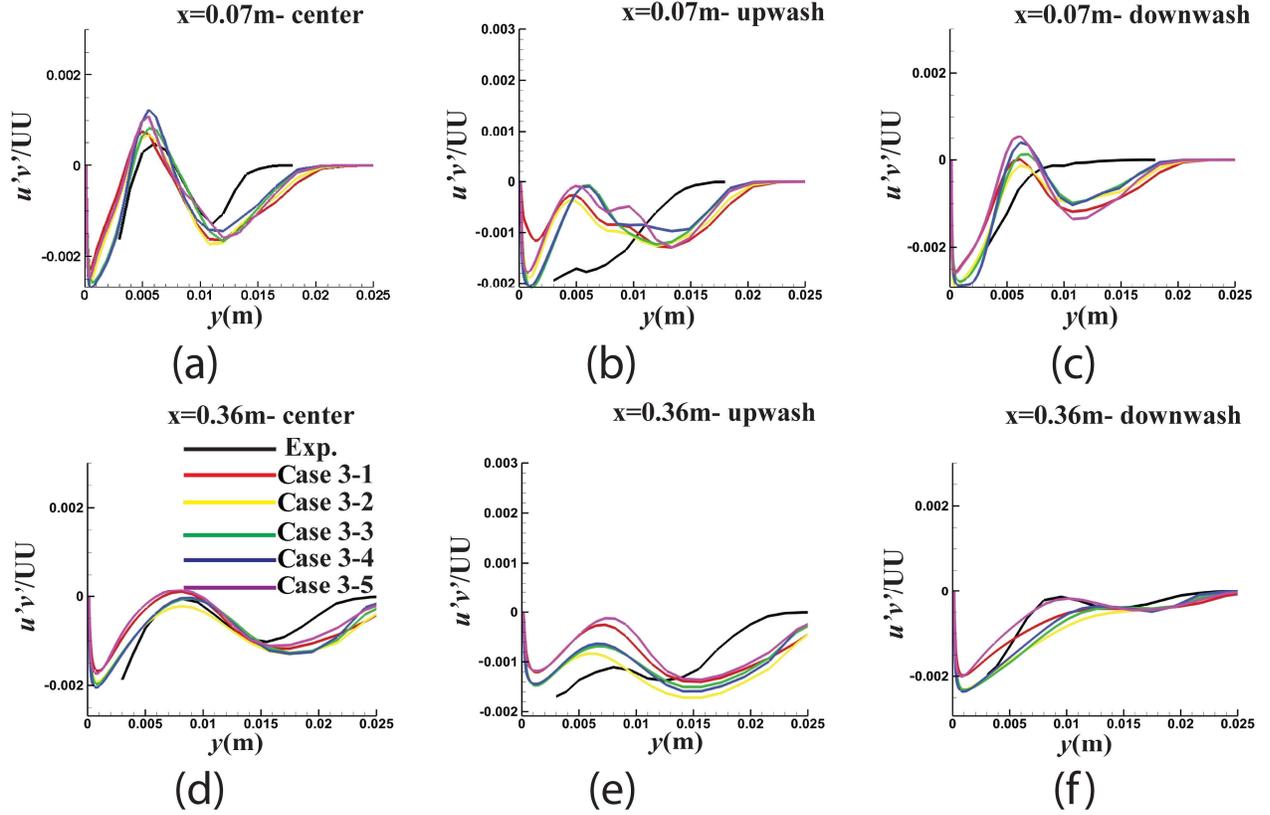

Figure 15. *u'v'* curves from experiment and different cases of two-equation $k$-$\omega$ model for (a) center, (b) upwash, and (c) downwash regions at $x = 0.07$ m cross-section, and for (d) center, (e) upwash, and (f) downwash regions at $x = 0.36$ m cross-section.

*3.4 Multi-Equation Turbulence Models*

The above shortcomings of the two-equation method necessitate considering multi-equation turbulence models, some of which are listed in Tab. 5. Fig. 16 shows velocity along the *x*-axis and indicates that contrary to the one- and two-equation methods, the multi-equation method has a stronger impact on the velocity calculation along this direction. Case 4-2 made the most accurate prediction of $u_x$. This model also outperformed the previous ones in terms of accuracy, especially in near-jet regions. The velocity calculation error for different cross-sections decreased to 5% on average compared with the earlier methods. Case 4-2 delivered better performance than the other methods and yielded results closer to experimental values. This model predicted $u_x$ in regions near to and far from the jet with 1% and 5% errors, respectively.

Table 5. Different cases of the multi-equation models.

| | Model details |
|---|---|
| Case 4-1 | Transition $k$-$k_L$-$\omega$ (three-equation) |
| Case 4-2 | Transition SST (four-equation), (options: production Kato-Launcher, production limiter) |
| Case 4-3 | Reynolds stress (seven-equation), (model: linear pressure-strain), (options: wall BC from $k$ equation, wall reflection effects), (near-wall treatment: standard wall functions) |
| Case 4-4 | Reynolds stress (seven-equation), (model: stress-$\omega$), (options: shear flow corrections) |
| Case 4-5 | Reynolds stress (seven-equation), (model: stress-BSL) |

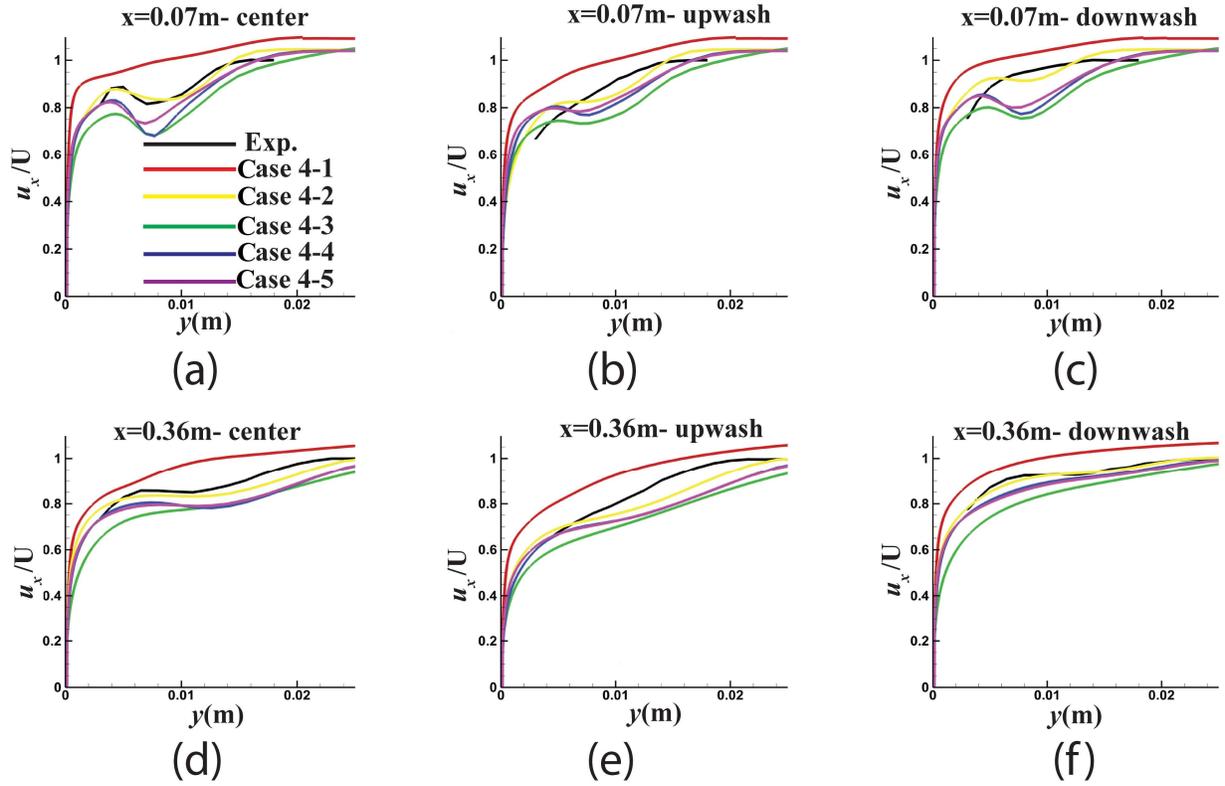

Figure 16. Velocity curves from experiment and different cases of multi-equation model for (a) center, (b) upwash, and (c) downwash regions at $x = 0.07$ m cross-section, and for (d) center, (e) upwash, and (f) downwash regions at $x = 0.36$ m cross-section.

Figs. 17 and 18 represent the results from multi-equation models used for predicting $u'u'$ and $u'v'$. It is observed that similar to Case 4-2, Cases 4-4 and 4-5 also made a better prediction of $u'u'$. However, Case 4-2 remained the superior model in terms of $u'v'$ prediction. There was a 5% error in the prediction of Reynolds stresses, especially at the center. It is worth noting that the model accuracy also improved in far-jet regions. Therefore, even though the results predicted by Cases 4-4 and 4-5 were similar to each other and relatively consistent with the experimental values, the transition SST model and its mentioned option (Case 4-2) seemed to yield more satisfactory results in the case of a fluid jet injected into the turbulent boundary layer in steady state conditions. In summary, the results concerning the multi-equation models indicated that Case 4-2 outperformed the other models in predicting both the velocity and the stress.

The steady state solution results discussed so far indicate that the one-equation SA method is sufficient as long as the only aim is a velocity analysis. If, however, an analysis of Reynolds stresses is also required, multi-equation models should come into play. Although two-equation methods are simple and efficient, it was observed that each of them yielded satisfactory results only in a particular portion of the flow and for particular parameters. Furthermore, the multi-equation models showed that using Case 4-2 was the best option for this problem in steady state conditions. Overall, this model could predict all of the parameters for different cross-sections with reasonable accuracy.

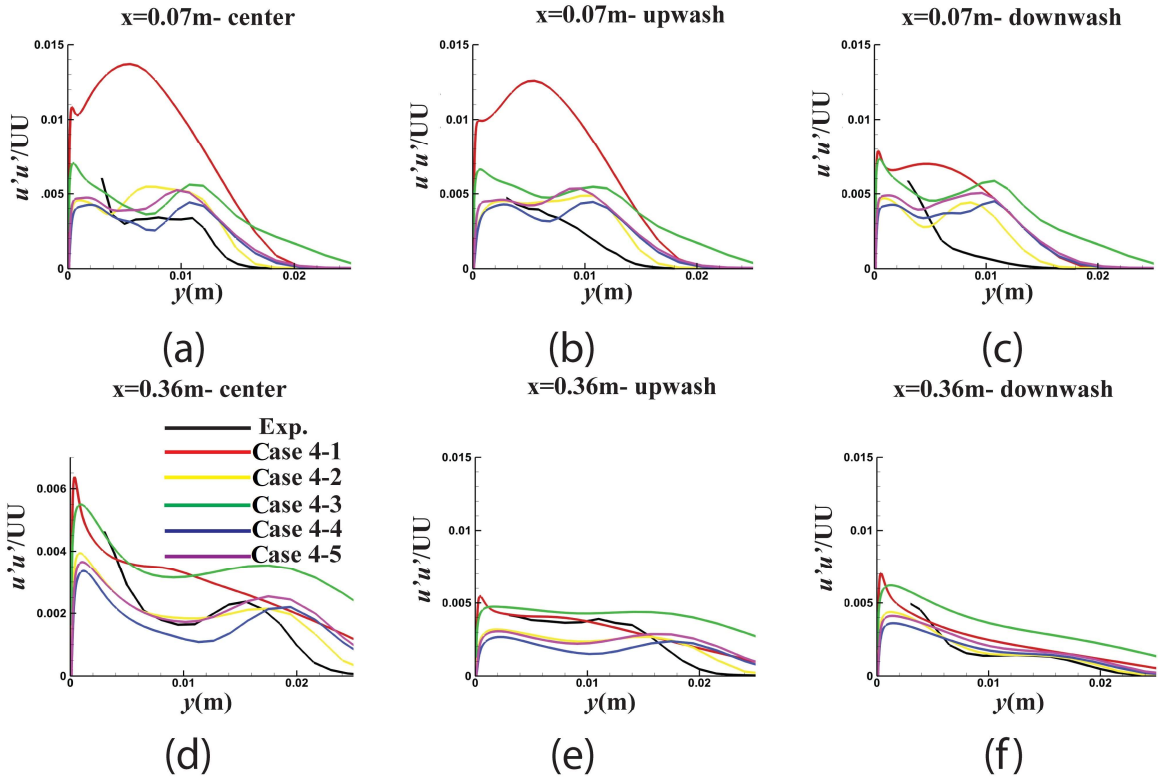

Figure 17. $u'u'$ curves from experiment and different cases of multi-equation model for (a) center, (b) upwash, and (c) downwash regions at $x = 0.07$ m cross-section, and for (d) center, (e) upwash, and (f) downwash regions at $x = 0.36$ m cross-section.

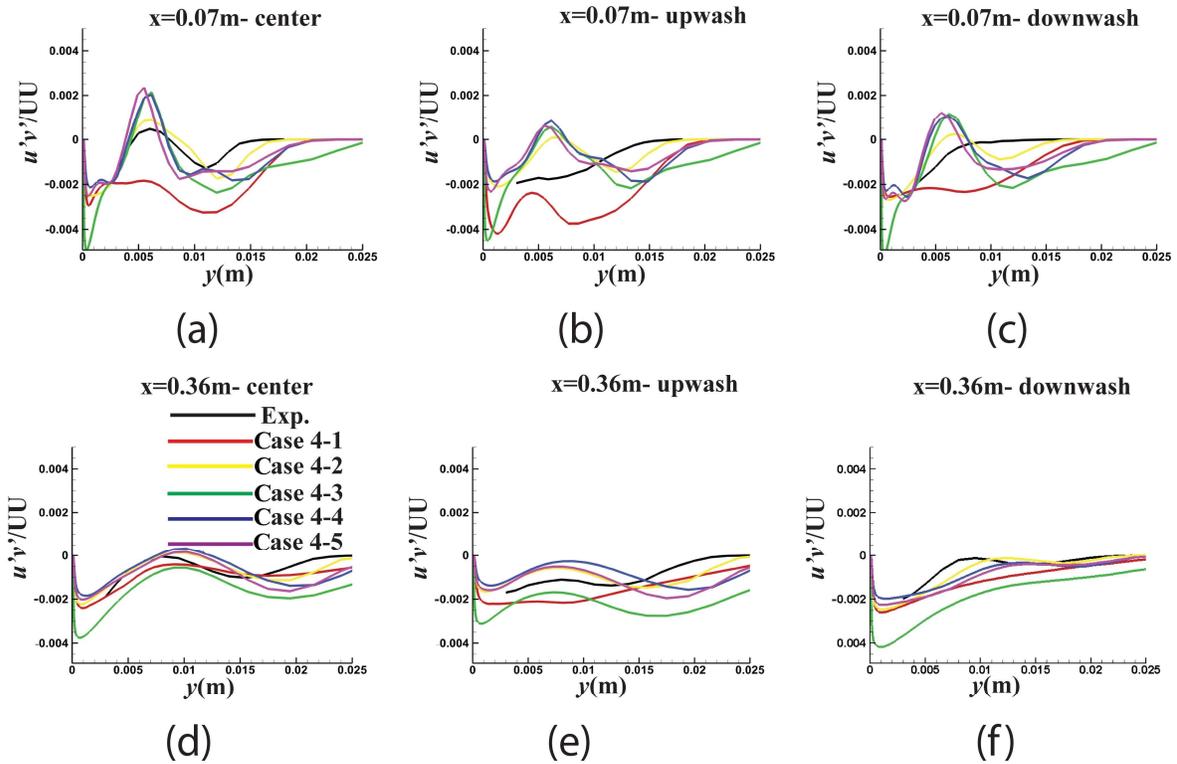

Figure 18. $u'v'$ curves from experiment and different cases of multi-equation model for (a) center, (b) upwash, and (c) downwash regions at $x = 0.07$ m cross-section, and for (d) center, (e) upwash, and (f) downwash regions at $x = 0.36$ m cross-section.

### 3.5 Transient Solution

Given the above discussion on the steady state solution, it is necessary to compare steady and transient methods and study their pros and cons. Tab. 6 lists some different models for transient analysis.

Table 6. Different considered cases for the transient state.

|  | Model details |
|---|---|
| Case 5-1 | Transition SST (four-equation), (options: production Kato-Launcher, production limiter) |
| Case 5-2 | Scale-adaptive simulation (SAS), (option: production limiter) |
| Case 5-3 | Scale-adaptive simulation (SAS), (options: production Kato-Launcher, production limiter, intermittency transition model) |
| Case 5-4 | Detached eddy simulation (DES), (RANS model: SST $k$-$\omega$), (DES option: delayed DES), (option: production limiter) |
| Case 5-5 | Detached eddy simulation (DES), (RANS model: realizable $k$-$\varepsilon$), (DES options: delayed DES) |

As the time step size can also affect the transient solution, first, Case 5-2 was used to analyze time step size independence. As shown in Fig. 19, this was done at two different locations close to and far from the jet within the boundary layer. The figure shows the velocity and Reynolds stress variations as the time step size functions. The results for different step sizes are observed to be almost the same when the solution reached a steady state, indicating that the steady solution did not substantially depend on the time step size. However, the step size is observed to have influenced the transient period. 0.0002 s was selected as the most suitable time step size to obtain valid results in both steady and transient states. According to the figure, a smaller step size would be computationally inefficient and unnecessary.

After selecting an adequate time step size, this section compares different models in terms of their capability of providing a transient solution. Fig. 20 shows the graphs of the transient state velocity in the *x*-direction. Cases 5-1 and 5-4 are observed to have yielded better results. As a transient state version of Case 4-2, Case 5-1 was the option of choice for predicting the velocity. Comparison of the results from Cases 5-1 and 4-2 reveals that the transient and steady state solutions were the same. We conclude that, in general, the results of the steady state solutions and those of the transient state were similar.

Figs. 21 and 22 represent the results for the transient state solution concerning the prediction of *u'u'* and *u'v'* by different models when they reached a steady state. The results indicate that each method outperformed the other ones in a different region. Cases 5-5 and 5-1 led to better *u'u'* predictions in the regions closer to and farther from the jet, respectively. Furthermore, the excellence of each of the different models in predicting *u'v'* is only for some cases. Therefore, no clear conclusions could be drawn about which models were superior. We can state as a general conclusion that, for the problem at hand, using steady state solution methods provided better results compared to transient state solution methods.

The DES model performed relatively poorly regarding computational cost, which can be attributed to requiring a finer grid. However, a finer grid would further increase the associated computational costs, rendering the method practically inefficient. The steady state solution was observed to be superior to the transient state solution. Moreover, a comparison of the time required for calculating the steady and transient state solutions by the same method shows that it took both methods more or less the same amount of time to obtain a solution. Therefore, Case 4-2 seemed to be generally superior in calculating all the parameters for different cross-sections.

Tab. 7 compares the above-mentioned cases by listing their pros and cons. This table can help for the selection of the best method by considering available resources and problem requirements. If the analysis aims to calculate the velocity without considering turbulence parameters, Case 1-1 is preferred to save time and reduce computational costs. On the other hand, if the analysis aims to calculate turbulence parameters, Case 4-2 is preferred. The advantages and disadvantages of each of the other models are provided in Tab. 7.

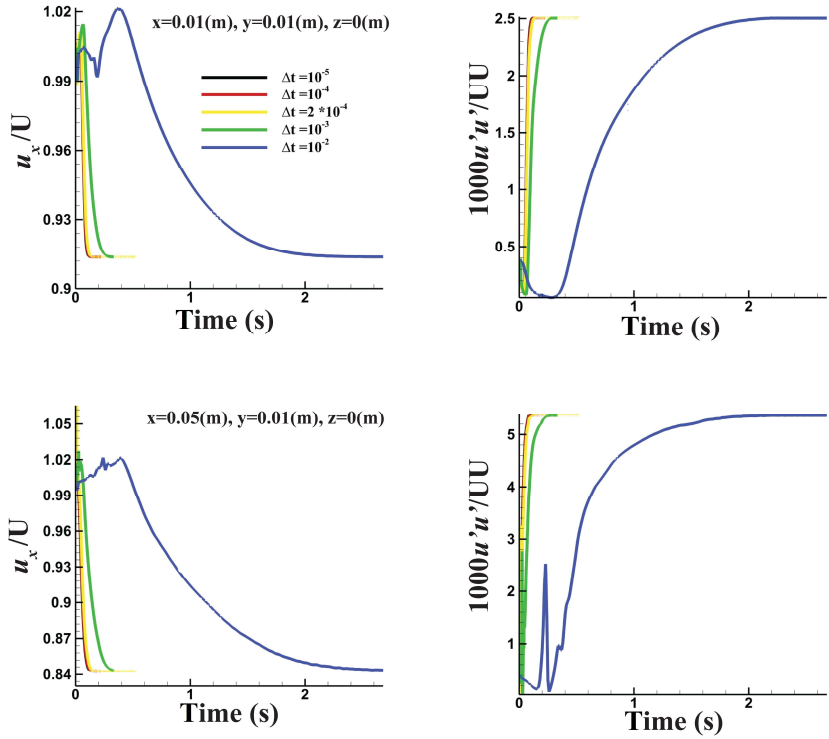

Figure 19. Time step size independence of the problem of microjet injection into a turbulent boundary layer.

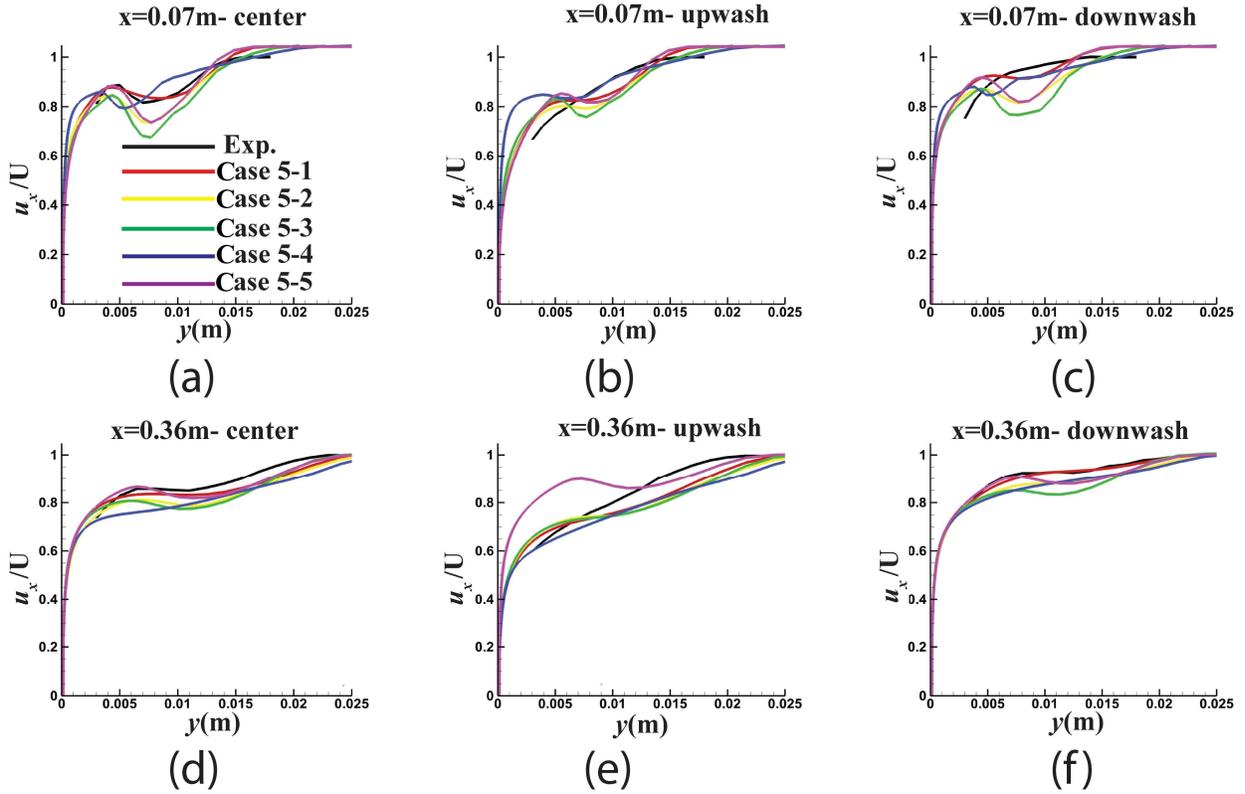

Figure 20. Transient solution velocity contours from experiment and different cases for (a) center, (b) upwash, and (c) downwash regions at $x = 0.07$ m cross-section, and for (d) center, (e) upwash, and (f) downwash regions at $x = 0.36$ m cross-section.

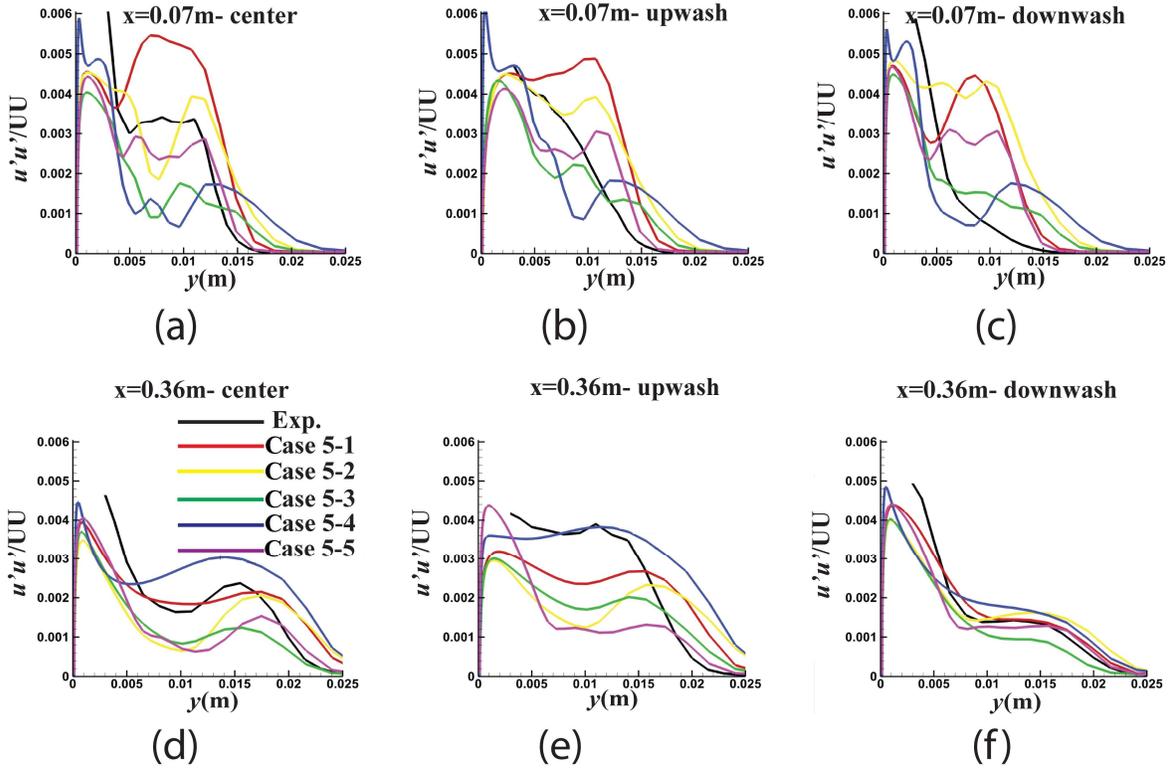

Figure 21. Transient state solution *u'u'* curves from experiment and different cases for (a) center, (b) upwash, and (c) downwash regions at $x = 0.07$ m cross-section, and for (d) center, (e) upwash, and (f) downwash regions at $x = 0.36$ m cross-section.

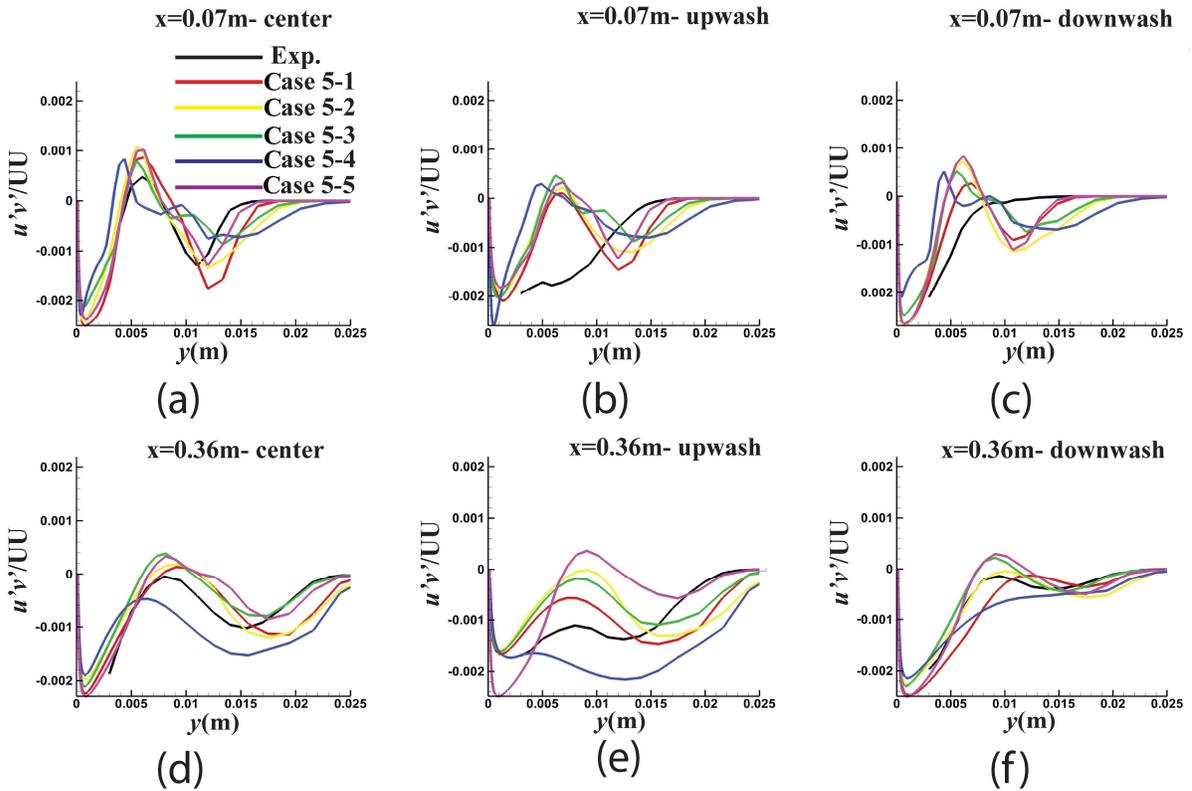

Figure 22. Transient state solution *u'v'* curves from experiment and different cases for (a) center, (b) upwash, and (c) downwash regions at $x = 0.07$ m cross-section, and for (d) center, (e) upwash, and (f) downwash regions at $x = 0.36$ m cross-section.

Table 7: Advantages and disadvantages of all cases.

|  | Turbulence model details | Advantages | Disadvantages |
|---|---|---|---|
| **Case 1-1** | Spalart-Allmaras (production: vorticity-based) | Low time and computational cost | Incapable of calculating parameters related to Reynolds stress |
| **Case 2-5** | $k$-$\varepsilon$ (model: standard), (near-wall treatment: Menter-Lechner) | Relatively low time and computational cost. Adequate calculation of $u'v'$ | Low accuracy in calculation of $u'u'$ |
| **Case 3-4** | $k$-$\omega$ (model: SST) (options: low-Reynolds-number corrections, production limiter) | Relatively low time and computational cost. Adequate calculation of $u'u'$ | Low accuracy in calculation of $u'v'$ |
| **Case 4-2** | Transition SST (four-equation) (options: production Kato-Launcher, production limiter) | Adequate accuracy in calculation of turbulence parameters | Relatively high time and computational cost |
| **Transient state solution** |  | Capability of examining graphs in the transient state mode | No special advantage in the accuracy of results compared to the steady state solution |

## 4. Conclusion

There are various prediction methods for solving turbulence flow problems, some of which might outperform the others for a particular problem. Therefore, analyzing and comparing the models is essential for any specific turbulence problem. This paper analyzed various turbulence models for predicting the behavior of a fluid jet injected into a turbulent boundary layer over a flat surface. The main findings are as follows:

- The one-equation SA model performs sufficiently well if the analysis is only aimed at velocity parameters. The results from this model indicated that it differed from the experimental results in regions far from and close to the jet by 10% and 3%, respectively.
- The two-equation $k$-$\varepsilon$ and $k$-$\omega$ models exhibited no advantage over the one-equation method. The difference between the experimental results and the results from the $k$-$\varepsilon$ model (Case 2-5) for $u'u'$ and $u'v'$ typically reached 40% and 10%, respectively. The corresponding value for the $k$-$\omega$ model (Case 3-4) was about 20% for both $u'u'$ and $u'v'$.
- Analysis of the multi-equation models revealed that Case 4-2 outperformed the other models in predicting the Reynolds stresses and velocity. The key advantage of this method was that its error in predicting Reynolds stresses dropped to 5%. This reduction in error was more pronounced at the center.
- According to the results from the transient state solution methods, these methods did not provide any significant advantage over the steady state solution methods. It was also observed that no one could be considered superior to other cases among the various cases in this group.

## Acknowledgments

This project is supported by National Natural Science Foundation of China (Grant No. 51575279).

**Nomenclature**----------------------------------------------

| | |
|---|---|
| $u$ | : Velocity in the $x$-direction |
| $v$ | : Velocity in the $y$-direction |
| $w$ | : Velocity in the $z$-direction |
| $U$ | : Inflow velocity |
| $u'$ | : Speed fluctuation in the $x$-direction |
| $v'$ | : Speed fluctuation in the $y$-direction |
| $w'$ | : Speed fluctuation in the $z$-direction |
| $V_{jet}$ | : Fluid microjet velocity |
| $\omega_x$ | : Vorticity in the $x$-direction |
| $Y^+$ | : A measure of application range of wall functions |
| $L$ | : Total surface length |
| $D$ | : Microjet diameter |
| $h$ | : Distance of the microjet from the center of the surface |
| $\alpha$ | : Jet pitch angle |
| $\beta$ | : Jet skew angle with respect to the x-axis |
| $\delta$ | : Boundary layer thickness |
| $\rho$ | : Fluid density |
| $\mu$ | : Fluid viscosity |
| $TI$ | : Turbulence intensity |
| $\mu_t/\mu$ | : Ratio of turbulent to molecular viscosity |
| $R$ | : Eddy-viscosity parameter |
| $\mathbf{b}_f$ | : Body force |

boundary layers." *Journal of Fluid Mechanics* 202 (1989): 263-294.